# A novel three-axis cylindrical hohlraum designed for inertial confinement fusion ignition

Longyu Kuang[1,2,3], Hang Li[1,2,3], Longfei Jing[1], Zhiwei Lin[1], Lu Zhang[1],

Lilin Li[1], Yongkun Ding[1,2,3,4,a)], Shaoen Jiang[1,2,3,4,b)],

Jie Liu[3,4,5], Jian Zheng[2,4]

1 Research Center of Laser Fusion, China Academy of Engineering Physics, P.O.Box 919-986, Mianyang 621900, China
2 CAS Key Laboratory of Basic Plasma Physics and Department of Modern Physics, University of Science and Technology of China, Hefei 230026, China
3 Center of Fusion Energy Science and Technology, Beijing 100088, China
4 Collaborative Innovation Center of IFSA, Shanghai Jiao Tong University, Shanghai 200240, China
5 Institute of Applied Physics and Computational Mathematics, Beijing 100088, China

a) E-mail: ding-yk@vip.sina.com.
b) E-mail: jiangshn@vip.sina.com.

## Abstract

A novel ignition hohlraum for indirect-drive inertial confinement fusion is proposed, which is named as three-axis cylindrical hohlraum (TACH). TACH is a kind of 6 laser entrance holes (LEHs) hohlraum, which is made of three cylindrical hohlraums orthogonally jointed. Laser beams are injected through every entrance hole with the same incident angle of 55°. The view-factor simulation result shows that the time-varying drive asymmetry of TACH is no more than 1.0% in the whole drive pulse period without any supplementary technology such as beam phasing etc. Its coupling efficiency of TACH is close to that of 6 LEHs spherical hohlraum with corresponding size. Its plasma-filling time is close to typical cylindrical ignition hohlraum. Its laser plasma interaction has as low backscattering as the outer cone of the cylindrical ignition hohlraum. Therefore, the proposed hohlraum provides a competitive candidate for ignition hohlraum.

## Introduction

In indirect-drive Inertial Confinement Fusion (ICF), laser beams inject into a high-Z hohlraum through laser entrance holes (LEHs) and are converted into X-ray radiation, then the radiation irradiates a low-Z capsule at the center of the hohlraum to bring the central fuel in the capsule to ignition conditions[1-6]. To achieve the ignition conditions, a convergence ratio above 40 is necessary in the central hot spot ignition scheme[1,3], so the drive asymmetry should be less than 1%[1], which is the critical point for hohlraum design. Up to now, cylindrical hohlraum is the main choice and was largely studied in the National Ignition Campaign (NIC)[7]. In order to obtain necessary time-varying symmetry in cylindrical hohlraums, multi-cone laser beams are used, and the $P_2$ and $P_4$ asymmetries are controlled by adjusting the power ratio between the inner and outer cones (beam phasing technology)[1,4]. However, the inner cone beams generate a considerable fraction of backscattering[7], and the overlapping of multiple cones near the LEHs cause crossed-beam energy transfer[8-10]. In addition, the plasma bubbles generated by outer cone affect the transfer of inner beams. These issues make the beam phasing a very complicated job. In addition, the beam phasing technology strictly depends on simulations, and the plasma of laser plasma interaction (LPI) region is non-local thermodynamic equilibrium, so it is difficult to be accurately calculated[11]. Besides cylindrical hohlraum, other hohlraums with different shapes have been proposed and investigated to improve the radiation environment inside the hohlraums, such as rugby hohlraum[12-14], 4 or 6 LEHs spherical hohlraum[15-20]. However, rugby hohlraums have the similar issue of beam phasing as in the cylindrical hohlraum. For 4 LEHs spherical hohlraum, it is difficult to suppress asymmetry below 1.0% using single cone beams[16,18]. 6 LEHs spherical hohlraum has the natural superiority in radiation symmetry[18], but experimental studies near ignition conditions are difficult and insufficiency on existing facilities.

## A novel three-axis cylindrical hohlraum

Based on plentiful studies on cylindrical hohlraum, a novel hohlraum named as three-axis cylindrical hohlraum (TACH) is proposed, which is made of three cylindrical hohlraums orthogonally jointed, as shown in Fig. 1. The axes of the three cylindrical hohlraums coincide with the X, Y, Z axis of a rectangular coordinate system respectively. TACH is a kind of 6 LEHs hohlraum. 8 laser quads arranged in one cone are injected through each LEH. The time-varying symmetry, coupling efficiency and plasma filling of TACH are studied in this article.

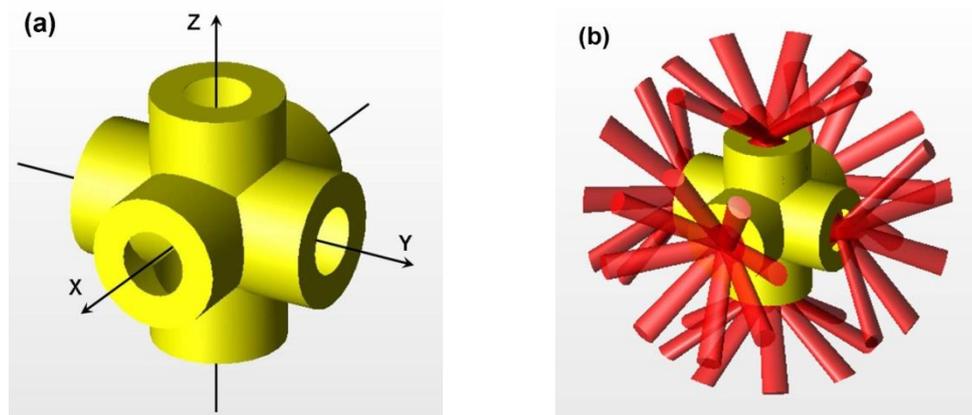

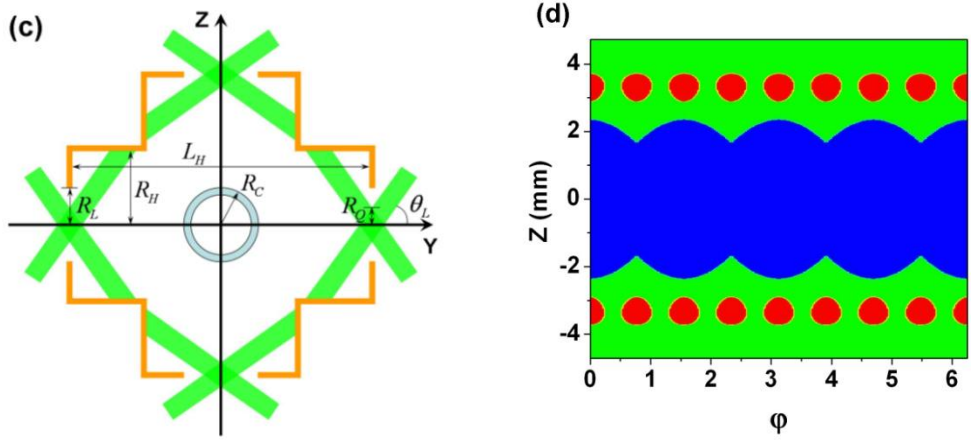

Fig. 1. Schematics of TACH. (a) 3D structure of TACH, (b) Laser arrangement of TACH, (c) Schematic of TACH on the Y-Z plane, (d) Unfolded inner wall of the cylindrical hohlraum using Z axis in TACH. Red region indicates laser spots on inner wall, green region indicates X-ray reemission region on inner wall, and blue region indicates inane region incised by the other two cylindrical hohlraums.

## Time-varying symmetry

Due to the single-cone design, the time-varying radiation drive symmetry of TACH won't be controlled by beam phasing technology as typical cylindrical hohlraum does. So it is crucial to study the time-varying symmetry of the single-cone TACH. The time-varying symmetry of TACH has been studied by use of a view-factor model[21-23]. In order to simplify the analysis, the radiation intensity of the laser spots and the reemission wall are assumed to be uniform distributed. $F_S$ and $F_W$ are used to denote the radiation intensity of the two kinds of regions. $T_r$ shown in Fig. 2(a) is chosen as the ignition radiation temperature pulse shape inside the hohlraum[4,24]. $\alpha_W$ and $\alpha_C$ are defined as the albedo of hohlraum wall and capsule respectively. $A_W$, $A_S$, $A_C$, and $A_L$ are used to denote the area of the hohlraum wall, laser spots, capsule surface and LEH respectively. Among them, $A_S$ is a part of $A_W$. According to the power balance equation[25],

$$F_S : F_W \approx 1 : \frac{(1-\alpha_W)A_W - A_L - (1-\alpha_C)A_C}{\alpha_W A_S} \quad (1)$$

A simple analytical model is used to calculate the time-varying albedo of the hohlraum wall. During 0~7 ns, the albedo is calculated by the scaling law $\alpha_W = 1 - 0.32 T_r^{-0.7} \tau^{-0.38}$ [2] and $\alpha_W$ is taken as 0.01 if $\alpha_W<0.01$. Then it increases linearly after 7 ns and reaches 0.9 at 13 ns [22,26] as shown in Fig.2 (a). The albedo calculated by the simple model is close to the simulation result from the 1D radiation hydrodynamic code of multi-groups (RDMG)[14], so it is reasonable to be used in our analysis.

With the ablation of x-ray and laser, the hohlraum wall moves inward. Study shows that the inward wall movement is dominated by re-radiated x-ray ablation, not by laser ablation[2]. Considering the quasi-uniform radiation environment inside the hohlraum, it can be assumed that the whole hohlraum wall including the LEH move inward at the same speed. Fig. 2(b) is a schematic of the hohlraum wall movement and the LEHs closure. According to the theory of radiation ablation[3], the speed ($v$) of hohlraum wall movement can be assumed to be proportional to the sound speed $c_s$ of the high-Z ablated plasma. For Au hohlraum wall, the sound speed is $c_s \approx 7.5 \times 10^4 T_r^{0.8} \rho^{-0.07}$ [1], where $T_r$ is the radiation temperature in eV, $\rho$ is the density of ablated plasma in g/cm$^2$. Ignoring the weak relationship between $c_s$ and $\rho$, it's clear that $v \sim T_r^{0.8}$. Assume the distance of inward wall motion is 275 μm at 14 ns[19,22,27]. According to the $T_r$ pulse shape shown in Fig. 2(a), the speed of hohlraum wall can be written as $v_{\mu m/ns} \approx 0.4 T_{r,eV}^{0.8}$. The displacement $s$ of hohlraum wall can be calculated by integrating the velocity over time as shown in Fig. 2(a).

As shown in Fig. 1(c), the laser quads are injected into the hohlraum with incident angle of $\theta_L$, the profile of which is set as round on the LEH plane with radius of $R_Q$. $R_H$, $R_L$ and $R_C$ are used to denote the radius of cylindrical hohlraum, LEH and capsule respectively. $\Delta$ indicates the minimum distance between the initial edge of the laser spots and the initial inner wall of adjacent cylindrical hohlraum as is shown in Fig. 2(b). So the length of the cylindrical hohlraum can be written as,

$$L_H \approx 2(R_H / \tan\theta_L + R_Q / \tan\theta_L + \Delta + R_H) \qquad (2)$$

In the calculation model, $R_Q$=0.6 mm, $\alpha_C$=0.3[20,28,29] and $R_C$=1.18 mm[4,24] are taken, which are all independent of time, and the initial $R_L$ is taken as 1.3 mm. By choosing the appropriate $\Delta$ to set the initial positions of laser spots and considering the time-varying albedo and displacement of hohlraum wall, the time-varying radiation symmetry on capsule $|\Delta F/\langle F \rangle|$ can be calculated, where F is the radiation intensity on capsule, $\Delta F \approx 0.5 (F_{max} - F_{min})$ and $\langle F \rangle$ is the average value of flux F on capsule[18].

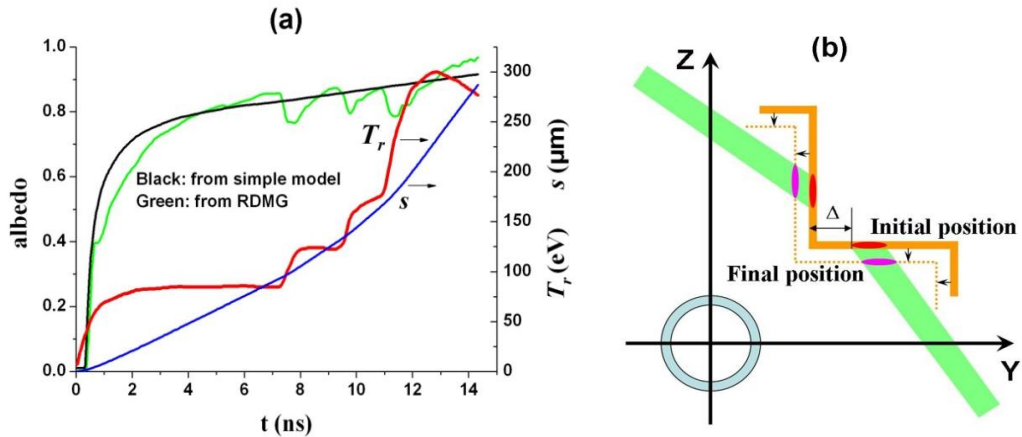

Fig. 2. (a) Evolution of the x-ray albedo from simple model (black) and RDMG simulation (green), the radiation temperature $T_r$ (red) and displacement $s$ of hohlraum wall (blue). (b) Schematic of the hohlraum wall motion and the LEHs closure. Solid golden line indicates initial position of the hohlraum wall, and dashed golden line indicates final position. Green strips indicate laser beams and their overlapping regions with hohlraum wall roughly indicate the locations of laser spots. The loop near the original point indicates the capsule. $\Delta$ indicates the minimum distance between the initial edge of the laser spots and the initial inner wall of adjacent cylindrical hohlraum.

In the calculation, $R_H/R_C$ is defined as the case-to-capsule ratio (CCR). The radiation drive symmetry of the hohlraums with CCRs=1.9, 2.0, 2.2 2.4 and 2.6 is investigated with a laser incident angle of 55 °. In the early stage, $\alpha_W \ll 1$, it can be calculated from Eq. (1) that the ratio of radiation intensity of laser spots to that of reemission wall $F_s/F_W \gg 1$. So the early symmetry is mainly determined by the initial position of the laser spots. In order to avoid the case that the laser beams from one cylindrical hohlraum enter another one, $\Delta \geq 0$ is taken as shown in Fig. 2(b). Calculations indicate that there is an optimum $\Delta$ in order to obtain the best symmetry for TACHs with different dimensions. Symmetry variation with $\Delta$ near the optimum $\Delta$ is illustrated in Fig. 3, and a smaller CCR corresponds to a better initial symmetry and a larger optimum $\Delta$.

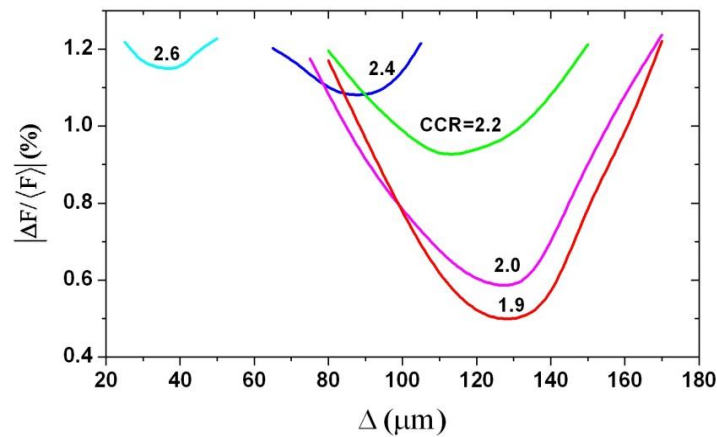

Fig. 3. Symmetry varies with $\Delta$ for hohlraums with different CCR near the optimum $\Delta$. A smaller CCR corresponds to a better initial symmetry and a larger optimum $\Delta$.

By setting the laser spots at the optimized initial position, the time-varying radiation drive symmetry of TACH is calculated by use of a view-factor model. The time-varying symmetry of TACH with different CCR is shown in Fig.4, which indicates that all the hohlraums with CCR from 1.9 to 2.6 have a good time-varying symmetry during the whole period of the laser pulse. The initial asymmetry increases with the CCR of the hohlraums. Nevertheless, even for the hohlraum with the greatest CCR of 2.6, the initial $|\Delta F/\langle F \rangle|$ is below 1.2%, and falls rapidly to below 1.0% after 300ps with the heating of the wall and the rising of the albedo. In the cases of CCR≤2.2, the initial $|\Delta F/\langle F \rangle|$ is less than 1.0%, meeting the need of the initial symmetry for ignition. During the period between 1ns and 11ns, all the $|\Delta F/\langle F \rangle|$ keep below 0.6%, the minimum value is even less than 0.2%. During the main pulse of the radiation (11~14ns), the asymmetry increases nearly linearly with time, a smaller CCR corresponds to a quicker increase of the asymmetry. At the end of the radiation pulse (t=14ns), the smaller asymmetry is achieved for hohlraums with larger CCR, which is just opposite from the early situations, and all of the asymmetry can be controlled below 1.0% at this time. After 1ns, the asymmetry changes slower for the hohlraum with a larger CCR, especially for the hohlraum with CCR=2.6 whose $|\Delta F/\langle F \rangle|$ keeps below 0.6% and almost independent of time from 1ns to 14ns.

In general, a smaller CCR corresponds to a better early symmetry while a larger CCR corresponds to a better final symmetry. Take the implosion of the capsule into consideration, a small initial CCR becomes larger and larger during the laser pulse, which is beneficial to improve the final symmetry of the TACH. So the TACHs with CCR=2.0~2.2 are selected as the optimum hohlraum.

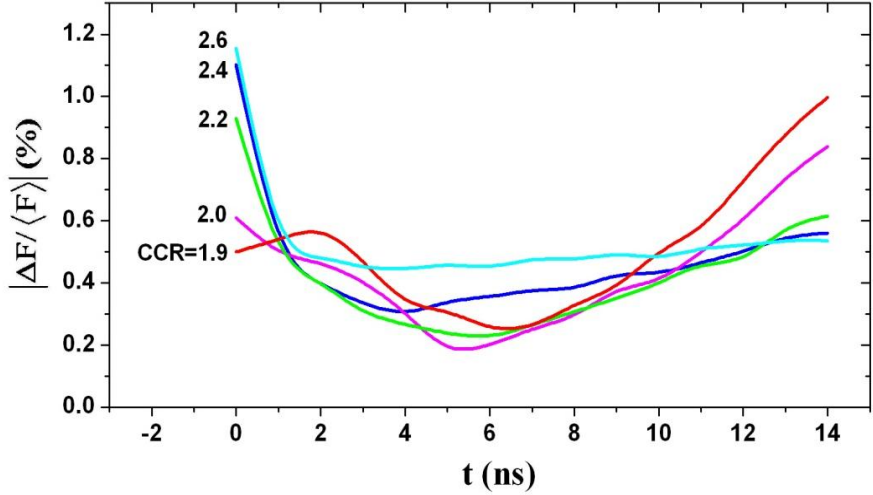

Fig. 4. Time-varying symmetry of the TACH with different CCR ($\theta_L$=55°).

Similar to the case of 6LEHs spherical hohlraum[18], as shown in the Fig. 5(a), there are two special types of points on the capsule. The normal directions of the first type points are parallel or antiparallel to the direction of X, Y or Z axis, and are facing the corresponding LEHs. One of such six points is marked with A in Fig. 5(a). For another type of points, the normal direction of which has an equal angle with X, Y and Z axis, adds up to 8, and one of such points is marked with B in Fig5(a). Calculations indicate that the initial symmetry on the capsule is optimal if the drive flux of A is equal to B. Then the relative flux of point A and its neighbourhood become more and more intensive than that of point B and its neighbourhood due to the movement of the laser spots towards the LEHs with time. At the end of the laser pulse, the radiation flux into the points facing the LEHs are more intensive than other points of the capsule.

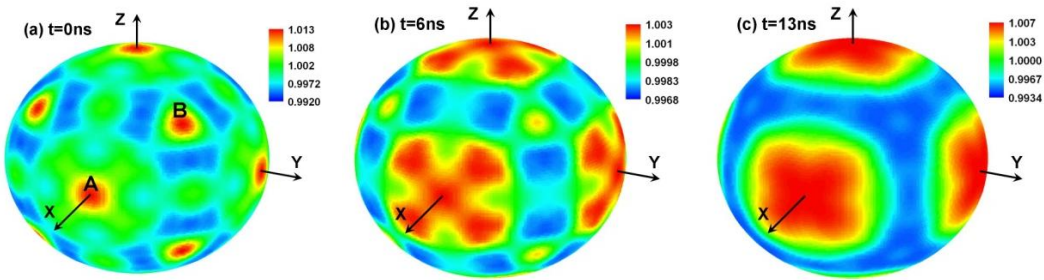

Fig. 5. Distribution of normalized drive intensity on the capsule at the moment of t=0 ns, 6 ns and 13 ns respectively (CCR=2.2, $\theta_L$=55°).

The time-varying symmetry varying with the incident angle of laser beams was also investigated under the conditions of CCR=2.2, as illustrated in Fig.6. The results show that the initial symmetry has weak relationship with the incident angle. However, this situation changes after 6ns. The time-varying symmetry becomes better with larger incident angle. The reason is that the distance of laser spots moving along the hohlraum axis is much smaller at a greater incident angle due to the inward movement of the hohlraum wall. Considering the trade-offs of time-varying symmetry and the laser injection convenience, 55 ° is chosen as the optimum incident angle of the laser beams.

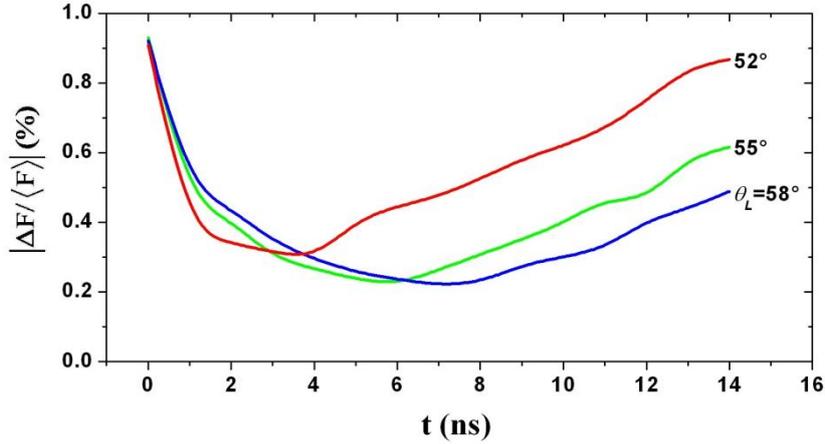

Fig.6 Time-varying symmetry varies with the incident angle of laser beams.

Based on the above analysis, the optimum size of TACH is CCR=2.0~2.2 and the optimum incident angle of laser is 55 °.

## Coupling efficiency

Define "coupling efficiency" as the ratio of absorbed energy $E_C$ by capsule to the incident laser energy $E_L$, which is given by

$$E_C / E_L \approx \eta_{aL}\eta_{LX}(1-\alpha_C)A_C / [(1-\alpha_W)A_W + (1-\alpha_C)A_C + A_L] \quad (3)$$

where $\eta_{aL}$ is the fraction of laser absorbed, and $\eta_{LX}$ is the laser to x-ray conversion efficiency.

The coupling efficiencies of several hohlraums are compared, including gas-filled cylindrical hohlraum (GFCH), near-vacuum cylindrical hohlraum (NVCH), six LEHs spherical hohlraum (SLSH) and TACH (CCR=2.0, 2.2). For the GFCH, $\eta_{aL}$ is about 0.85 due to the strong backscatter from the inner cone beams[7]. For the NVCH, $\eta_{aL}$ is about 0.96[30]. For SLSH and TACH, $\eta_{aL}$ is about 0.96, because the properties of LPI of SLSH and TACH are similar with the outer cone of GFCH with a low level of backscattering. $\eta_{LX}$ is about 0.8 for all kinds of hohlraums[31]. Time-varying coupling efficiency comparison of GFCH ($R_H$=2.95mm, $L$=10.6mm, LEH⌀3.1mm)[4,24], NVCH ($R_H$=3.36mm, $L$=11.26mm, LEH⌀3.9mm)[32], SLSH (CCR=4.0, 4.4, LEH⌀2.6mm) and TACH (CCR=2.0, 2.2, LEH⌀2.6mm) are calculated and shown in Fig.7. The calculation results show that the time-varying coupling efficiency of TACH with CCR0 is very close to that of SLSH with 2×CCR0. The coupling efficiencies of TACH (CCR=2.0), SLSH (CCR=4.0) and NVCH are close to each other, which is about 13% lower than that of GFCH. The coupling efficiency of TACH (CCR=2.2) is similar with SLSH (CCR=4.4), which is about 20% lower than that of GFCH. Nevertheless, it is worthwhile to spend 13%~20% more laser energy for a higher symmetry during the whole period of implosion of capsule. Furthermore, the coupling efficiency of TACH can be increased by above 10% using LEH shields[20]. Compared with the arrangement of multi-cone lasers for cylindrical hohlraum, the arrangement of single-cone lasers with large angle can supply sufficient space to place LEH shields.

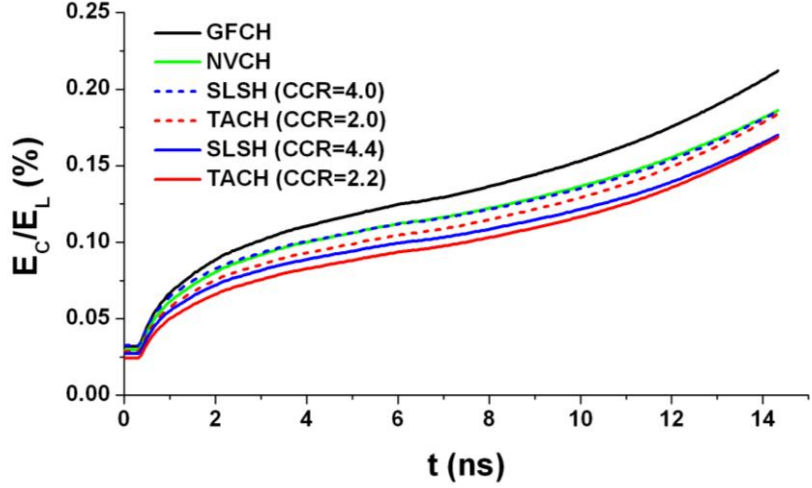

Fig.7 Time-varying coupling efficiency comparison of GFCH ($R_H$=2.95mm, $L$=10.6mm, LEH$\varnothing$3.1mm), NVCH ($R_H$=3.36mm, $L$=11.26mm, LEH$\varnothing$3.9mm), SLSH (CCR=4.0, or CCR=4.4, LEH$\varnothing$2.6mm) and TACH (CCR=2.0, or CCR=2.2, LEH$\varnothing$2.6mm). Capsules with same $R_C$=1.18mm are adopted.

## Plasma filling

The plasma ablated from the hohlraum wall will fill the volume inside the hohlraum, which would affect the injection of laser and limit the performance of hohlraum. The plasma filling time of TACH and cylindrical hohlraum are compared to evaluate the plasma filling issue of TACH. Filling model in Ref. 33 is only used in cylindrical hohlraums, which is extended to other hohlraums with arbitrary shape in this article. $V_H$ is defined as the hohlraum volume, $V_{fill}$ is the residual hohlraum volume deducting the volume of laser channels and capsule from $V_H$, so the radiation-ablated plasma is filled in $V_{fill}$. Define $\beta=A_W/V_{fill}$. Laser wavelength is 351nm. Unit selection: laser power $P_0$ uses TW, radiation temperature $T_r$ uses eV, electron temperature $T_e$ uses keV, length uses cm, $n_i$ and $n_e$ uses $n_c$, and filling time $\tau$ uses ns.

$A_{loss}$ is defined as the equivalent energy loss area,

$$A_{loss} \ll (1 \quad \alpha_W)A_W \quad (1 \quad \alpha_C)A_C \quad A_L \qquad (4)$$

According to the power balance of hohlraum,

$$P_0 \ll 5.67 \times 10^{-8}(\eta_{aL}\eta_{LX})^{-1} A_{loss} T_r^4 \tag{5}$$

According to the power balance of laser channel[33],

$$P_0 \ll 0.159 T_e^5 \sqrt{1-n_e}/(Z_h^2 n_e^2) \tag{6}$$

According to the pressure balance between laser channel plasma and radiation-ablated plasma,

$$1000 n_e T_e \ll Z n_i T_r \tag{7}$$

Using Eqs. (128) and (129) in Ref. 1 to calculate radiation-ablated mass rate, and ablated plasma from the hohlraum wall $A_W$ is supposed to uniformly fill in residual hohlraum volume $V_{fill}$. Then,

$$n_i \ll 6.57 \times 10^{-8} T_r^{1.86} \tau^{0.75} \beta \tag{8}$$

The ionization degree of background plasma calculated from Eq. (187) in Ref. 1,

$$Z \ll 2.9 T_r^{0.45} \tag{9}$$

The ionization degree of plasma in the laser channel[33],

$$Z_h \ll 40 T_e^{0.2} \tag{10}$$

From Eqs. (4) to (10),

$$\tau \ll 1.06 \times 10^{12} T_r^{-3.26} n_e^{1.91}(1-n_e)^{-0.145}(\eta_{aL}\eta_{LX})^{-0.29} A_{loss}^{0.29} \beta^{-1.33} \tag{11}$$

Eq. (11) can be used to calculate the filling time of hohlraums with arbitrary shape. From Eq. (11), for a certain $T_r$, $n_e$, $\eta_{aL}$ and $\eta_{LX}$, plasma filling time

$$\tau \frown A_{loss}^{0.29} \beta^{-1.33} \tag{12}$$

Define ε as the ratio of the filling time of TACH to the filling time of cylindrical hohlraum. Assume that $V_{fill}$ is equal to $V_H$. Take $\alpha_W$=0.8, $\alpha_C$=0.3, $R_C$=1.18mm for two kinds of hohlraum. For TACH (Δ=100μm, LEHø2.6mm) and cylindrical hohlraum ($R_H$=2.95mm, $L$=10.6mm, LEHø3.1mm), ε can be calculated from Eq. (12). Fig.8 shows the ε variation with CCR of TACH. For TACHs with CCR between 2.0 and 2.2, ε is between 0.96 and 1.1, so the filling time of TACH is close to cylindrical hohlraum.

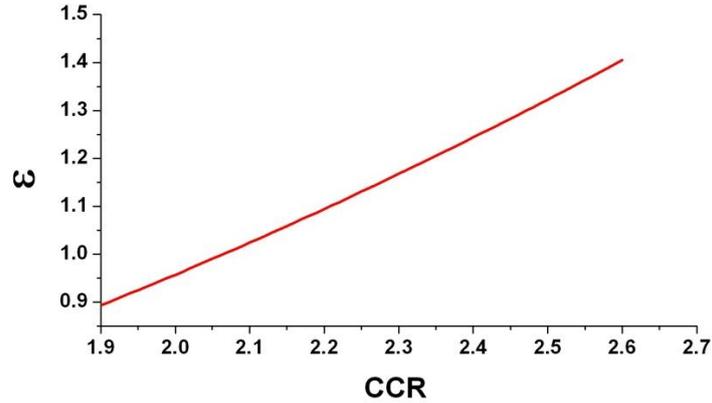

Fig.8 Ratio of filling time of TACH (Δ=100μm, LEHø2.6mm) to filling time of cylindrical hohlraum ($R_H$=2.95mm, $L$=10.6mm, LEHø3.1mm) varying with CCR.

## Discussion

From another point of view, TACH is composed of six half cylindrical hohlraums (HCHs). For laser injection, the six HCHs can be decoupled from each other, which bring great convenience for laser arrangement. Furthermore, the lasers and plasma condition in each HCH are mainly cylindrical symmetry, so these can be studied approximately by a cylindrical 2D radiation hydrodynamic model. Single cone lasers are injected into each HCH with large incident angle, which is similar to the outer cone of the ignition cylindrical hohlraum in NIC. Therefore, it is reasonable to predict the backscattering of TACH is as slight as that of the outer cone of the ignition cylindrical hohlraum. In addition, single-cone injection avoid several other LPI problems of multi-cone cylindrical hohlraums, such as crossed-beam energy transfer between two laser cones, and blocking of the transfer of inner-cone laser by high-Z plasma bubbles ablated by outer-cone lasers. Single-cone injection greatly simplifies the symmetry tuning. To optimize the symmetry of a certain TACH, it is only need to adjust the initial position of laser spot and control the power balance of laser. The parameters of TACH can be optimized to control the time-varying asymmetry below 1.0% during the whole drive pulse. The filling time of TACH is close to typical ignition cylindrical hohlraum in NIC, but the coupling efficiency of TACH is about 13%~20% lower than that of ignition cylindrical hohlraum. Nevertheless, it is worthwhile to spend 13%~20% more laser energy for the superiorities of TACH as discussed above.

## Summary

In summary, a novel orthogonal three-cylinder hohlraum is proposed for inertial confinement fusion ignition. The time-varying symmetry, coupling efficiency and plasma filling time of this hohlraum are investigated in this article. Study shows that the optimized TACH can control the time-varying asymmetry below 1.0% during the whole drive pulse. Meanwhile, the symmetry tuning is greatly simplified and the risks of LPI are greatly reduced compared with typical ignition cylindrical hohlraum as used in NIC. The extended filling model shows that the plasma filling time of TACH is close to cylindrical hohlraum. Although 13%~20% more energy is needed for TACH to generate the same radiation temperature as cylindrical hohlraum, it is worthwhile for the benefits of TACH. Therefore, the proposed hohlraum provides an important potential way for hohlraum design in ICF.

## Acknowledgments

The authors wish to acknowledge the beneficial discussions with Dr. Dong Yang. This work was supported by the National Natural Science Foundation of China (Grand Nos. 11435011, 11475154, 11305160, 11405160, 11505170), the Found of National Science and Technology on Plasma Physics Laboratory (No. 9140C680104140C68287), and the Found of Center of Fusion Energy Science and Technology (No. J2014-0401-04).